\documentclass[
reprint,
amsmath,amssymb, 
prl, 
jmp,
floatfix,
superscriptaddress,
]{revtex4-2}

\usepackage{physics}
\usepackage{graphicx}
\usepackage{xfrac}
\usepackage{bm}
\usepackage[hidelinks]{hyperref}
\usepackage{nameref}
\usepackage{mathrsfs}
\usepackage{xcolor}
\DeclareFontFamily{OT1}{pzc}{}
\DeclareFontShape{OT1}{pzc}{m}{it}{<-> s * [1.10] pzcmi7t}{}
\DeclareMathAlphabet{\mathpzc}{OT1}{pzc}{m}{it}
\usepackage[mathlines]{lineno}%

\newcommand{\mpi}{Max-Planck-Institut f{\"u}r Kernphysik, Saupfercheckweg 1, 69117 Heidelberg, Germany.}
\newcommand{\itp}{Institut f{\"u}r theoretische Physik, Ruprecht-Karls-Universit{\"a}t Heidelberg, Philosophenweg 19, 69120 Heidelberg, Germany.}

\begin{document}

\title{Resolving Vibrations in a Polyatomic Molecule with Femtometer Precision}

\author{Patrick Rupprecht}
\email{patrick.rupprecht@mpi-hd.mpg.de}
\affiliation{\mpi}

\author{Lennart Aufleger}
\affiliation{\mpi}

\author{Simon Heinze}
\affiliation{\itp}

\author{Alexander Magunia}
\affiliation{\mpi}

\author{Thomas Ding}
\affiliation{\mpi}

\author{Marc Rebholz}
\affiliation{\mpi}

\author{Stefano Amberg}
\affiliation{\mpi}

\author{Nikola Mollov}
\affiliation{\mpi}

\author{Felix Henrich}
\affiliation{\mpi}

\author{Maurits W. Haverkort}
\affiliation{\itp}

\author{Christian Ott}
\email{christian.ott@mpi-hd.mpg.de}
\affiliation{\mpi}

\author{Thomas Pfeifer}
\altaffiliation[Also at ]{Center for Quantum Dynamics, Ruprecht-Karls-Universit{\"a}t Heidelberg, Im Neuenheimer Feld 226, 69120 Heidelberg, Germany.}
\email{thomas.pfeifer@mpi-hd.mpg.de}
\affiliation{\mpi}

\date{\today}

\begin{abstract}
We measure molecular vibrations with femtometer precision using time-resolved x-ray absorption spectroscopy. For a demonstration, a Raman process excites the A$_{1g}$ mode in gas-phase SF$_6$ molecules with an amplitude of $\approx50$\,fm, which is probed by a time-delayed soft x-ray pulse at the sulfur $L_{2,3}$-edge. 
Mapping the extremely small measured energy shifts to internuclear distances requires an understanding of the electronic contributions provided by a many-body \textit{ab initio} simulation. 
Our study establishes core-level spectroscopy as a precision tool for time-dependent molecular-structure metrology. 

\end{abstract}

\maketitle

\begin{figure}
    \centering
    \includegraphics[width=0.48\textwidth]{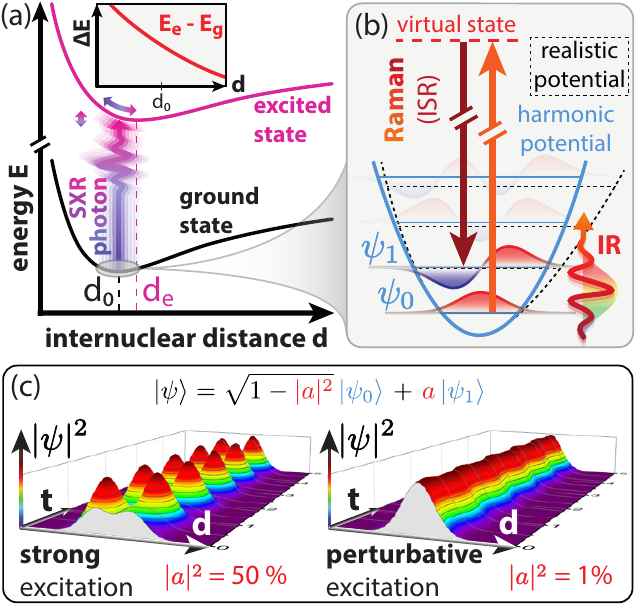}
    \caption{\textbf{Mapping vibrational dynamics with core-level spectroscopy.}
    (a) Exemplary PECs of a molecular electronic ground and excited state which can be resonantly coupled by SXR radiation. Depending on the internuclear distance $d$, the energy differences of the PECs translate into different transition energies $\Delta E$, which is illustrated around the equilibrium internuclear distance $d_0$ in the inset. 
    (b) Vibrational excitation scheme: The ground and first excited nuclear wave functions, $\psi_0$ and $\psi_1$, of the electronic ground-state PEC are illustrated. 
    A broadband, ultrashort IR pulse can populate the excited state via ISR in a two-photon process. 
    (c) Temporal evolution of the resulting vibrational wave packet $|\psi (t)|^2$ over five oscillation periods for $|a|^2$~=~50\% population (left) of $\ket{\psi_1}$ and for perturbative excitation ($|a|^2$~=~1\%, right).
    }
    \label{fig1}
\end{figure}

\begin{figure}[b]
    \centering
    \includegraphics[width=0.43\textwidth]{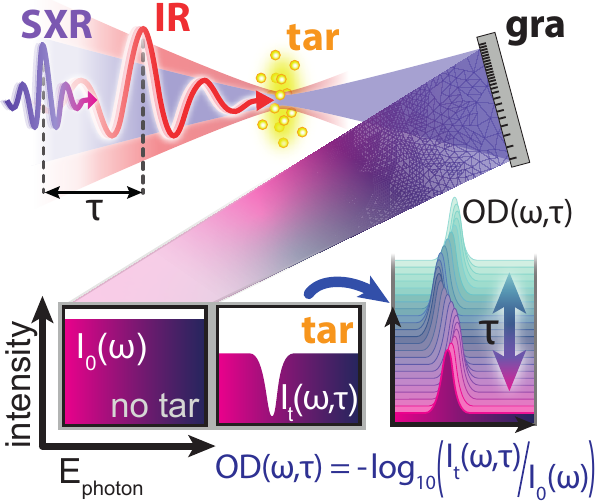}
    \caption{\textbf{Experimental TR-XAS scheme.}
    A co-propaga-ting weak SXR and strong IR pulse are time-delayed with respect to each other by $\tau$. They are focused into a molecular target (tar), where the preceeding IR pulse initiates vibrational dynamics [see Fig.~1(b)] while the SXR pulse probes the system  [cf. Fig.~1(a)]. The transmitted SXR pulse is spectrally dispersed by a grating (gra) and measured with an SXR-sensitive CCD camera. Using a reference spectrum without target, $I_0(\omega)$, and the transmitted spectra at different times $\tau$, $I_t(\omega, \tau)$, a 2D-OD($\omega, \tau$) map encodes signatures of electronic and vibrational dynamics.
    }
    \label{fig2}
\end{figure}

Molecular physics and chemistry are governed by electron dynamics but ultimately realized in the structural coordination of the nuclei. The most subtle molecular structure alterations are vibrations.
Hence, exciting and precisely measuring molecular vibrations is crucial for understanding chemical reactions and their control \cite{zavriyev1993light,crim1996bond,letokhov1973use,cederbaum1983multistate,zewail2000femtochemistry,weinacht2001coherent,neugebauer2004vibronic}. 
Especially the interplay between and dynamics of electrons and nuclei are of pivotal interest \cite{fulton1961vibronic,cederbaum1983multistate,worth2004beyond,kobayashi2019direct,zinchenko2021sub}. One frontier is the single quantum level by investigating smallest electronic or vibrational \cite{cerchiari2021measuring,velez2019preparation} excitations.\\
\indent Lasers are a powerful tool for controlling molecular dynamics. In the last decades, electronic population \cite{vitanov2017stimulated}, molecular rotation \cite{karamatskos2019molecular} and highly excited vibrational states \cite{baumert1991femtosecond,wustelt2021laser} have been successfully targeted by intense light fields. Laser-based infrared (IR) and Raman spectroscopy are well established techniques in science and industry \cite{lewis2001handbook} for investigating electronic and vibrational molecular characteristics in matter. In combination with core-level spectroscopy \cite{siegbahn1982electron} one gains deep insights into the equilibrium electronic and vibrational structures of molecules and solids \cite{ament2011resonant}.\\
\indent On the other hand, diffraction methods are commonly used to determine the distances and dynamics within matter using, e.g., 
electrons \cite{sciaini2011femtosecond,shen2019femtosecond,yang2016diffractive,champenois2021conformer,zewail2010four, blaga2012imaging,wolter2016ultrafast,sanchez2021molecular}, 
neutrons \cite{jacrot1976study} 
or photons \cite{warren1990x, kupper2014x}.
For gas-phase molecules, these diffraction methods provide nano- to picometer spatial resolutions.
Focusing on molecular vibrations, many spectroscopy experiments have revealed dynamics in a time-resolved manner \cite{wagner2006monitoring,kunnus2020vibrational,hosler2013characterization,timmers2019disentangling,kobayashi2020coherent,saito2019real,zinchenko2021sub,pertot2017time}. 
So far, the smallest spatially resolved vibrational feature has been measured at 0.6\,pm \cite{wei2017elucidating}. 
A quantification of structural changes on the few-femtometer level \cite{pettifer2005measurement,weisshaupt2017ultrafast,jo2018measuring,de2021femtometer} and potentially below \cite{kozina2014measurement,wu2020graphene} has been thus far limited to solid-state systems.\\
\indent In this work, we demonstrate time-resolved vibrational metrology of neutral, gas-phase molecules with an unprecedented 14\,fm spatial precision. For a proof-of-principle experiment, we conduct measurements of smallest bond-length changes in sulfur hexafluoride (SF$_6$) using table-top, time-resolved x-ray absorption spectroscopy (TR-XAS). Here, sulfur (S) $L_{2,3}$-edge transitions 
around 173\,eV are probed. 
With this all-optical spectroscopic method, we induce and time-resolve coherent vibrational excitations of a molecular ensemble in the perturbative limit.\\
\indent The general mechanism that enables the tracing of extremely small structural changes with TR-XAS is illustrated in Fig.~\ref{fig1}(a): Since TR-XAS probes the dipole response of the molecular system, the potential-energy curves (PECs) of the involved electronic states are crucial. Incoming soft x-ray (SXR) radiation induces a dipole transition from the electronic ground-state PEC to an excited state. 
These two PECs may generally differ in their minimum-energy position ($d_0$ and $d_e$) as well as in their shape and width. 
Ultrashort SXR pulses hence can be used to map the internuclear distance $d$ to the PEC energy difference $\Delta E$ [see inset in Fig.~\ref{fig1}(a)]. 
For sufficiently small internuclear changes $\Delta d$, this mechanism is general and can be applied independently of the excited-state-PEC character. 
Thus, very small alterations of the electronic ground-state nuclear wave packet can be detected. One widely applicable process to induce such perturbations in the probability-density distribution within the ground-state PEC is nonresonant, impulsive stimulated Raman (ISR) excitation \cite{walsh1989theory,weiner1990femtosecond,wittmann2000fs,wagner2006monitoring} via an ultrashort IR pulse. 
Fig.~\ref{fig1}(b) illustrates how the first vibrationally excited state is coherently populated by coupling it to the ground state in the two-photon ISR process via a virtual state. The ground- and excited-state nuclear wave functions, $\psi_0$ and $\psi_1$ with respective energies $\hbar \omega_0$ and $\hbar \omega_1$, can be well approximated by the respective ones from a harmonic potential \cite{wilson1980molecular}.
The temporal evolution of the vibrational superposition state $\ket{\psi (t)} = \sqrt{1 - |a|^2} \ e^{-i\, \omega_0 t}\ket{\psi_0} + a \ e^{-i\, \omega_1 t}\ket{\psi_1}$ after excitation results in a time-dependent probability density $\left|\psi(t)\right|^2$ for different relative populations $|a|^2$ of the first excited vibrational state $\ket{\psi_1}$ [see Fig.~1(c)]. While a strong excitation of e.g., $|a|^2 = 50$\% leads to a quantum-hopping behavior (Fig.~\ref{fig1}(c) left), a perturbative excitation (e.g., $|a|^2 = 1$\% in Fig.~\ref{fig1}(c) right) approximates an internuclear-distance oscillation of the ground-state-like wave packet. The question arises as to what minimal coherent vibrational excitation and thus bond-length change is measurable within an ensemble of molecules.\\
\indent To shed light on this extreme perturbative case of coherent molecular excitation, a table-top TR-XAS experimental scheme as outlined in Fig.~\ref{fig2} was employed. The optical setup delivers IR laser pulses (center wavelength $\lambda_c$~=~1535\,nm) with 1\,mJ pulse energy and a measured three-optical-cycle duration ($\tau_{FWHM}$~=~15\,fs) at a 1\,kHz repetition rate. Focusing these IR pulses into a neon-filled cell inside a vacuum beamline generates SXR photon energies up to 200\,eV in a high-order harmonic generation process. Afterwards, IR and SXR are spatially separated and time-delayed with respect to each other. 
Finally, the strong IR 
and comparatively weak SXR beams are refocused into an effusive cell (interaction length 3~mm) 
which is filled with 16\,mbar of SF$_6$. The transmitted SXR spectrum is dispersed by a grating and measured with a CCD camera. Finally, the measured data are evaluated in terms of spectral absorbance (optical density; OD). 
Varying the SXR-IR time delay $\tau$ results in a time-dependent absorbance OD($\omega, \tau$). \\
To evaluate the data, we focused on the S $L_{2,3}$-edge absorbance doublet 6a$_{1g}$. 
This resonance is linked to a dipole transition from the spin-orbit-split, core-level S~$2p$ orbital \mbox{2t$_{1u}\,(j_{\pm} = 3/2,\,1/2$)} to the lowest unoccupied molecular orbital (LUMO) 6a$_{1g}$ \cite{hudson1993high,dehmer1972evidence}. The time-delay-dependent resonant absorbance spectra for an IR intensity of \mbox{I$_{IR,1}=9$\,TW/cm$^2$} are shown in Fig.~3(a). An example of a single measured 2t$_{1u}\!\!\!\rightarrow\,$6a$_{1g}$  doublet absorbance is given in Fig.~3(b). To extract small time-dependent signatures, a mean over SXR-first spectra 
was subtracted from all OD($\tau$) absorbances. The resulting $\Delta$OD($\omega, \tau$) in Fig.~3(c) shows an oscillatory behavior in the IR-first region ($\tau >$~0\,fs). For quantification purposes, a Voigt model was fitted to the OD($\omega, \tau$) data from Fig.~3(a). A comparison of the experimental $\Delta$OD [Fig.~3(c)] and the $\Delta$OD based on the fits [Fig.~3(d)] verifies that the measured time-dependent features are reproduced by the fits. 
The periodicity of the oscillatory signatures in the data coincides with the symmetric breathing mode vibration A$_{1g}$ of SF$_6$ with a 43\,fs period \cite{aboumajd1979analysis,wagner2006monitoring}. Such a vibration is visualized in Fig.~3(e). 
In addition, a second measurement was conducted with a higher IR intensity of I$_{IR,2}=26$\,TW/cm$^{2}$ 
which allows to investigate the IR-intensity dependence of the involved electronic and vibrational dynamics. To summarize the fit results of both data sets, Fig.~3(f) shows the resonance energy shift $\Delta E_{c}$ for both doublet peaks 6a$_{1g}(j_{\pm})$ and both intensities (I$_{IR,1/2}$) matched to each other (right ordinate axis for I$_{IR,2}$ data set). \\
\indent In order to gain further insights into the oscillatory behavior of the 6a$_{1g}$ peak energies for the IR-first case, a quantum-mechanical \textit{ab initio} simulation was combined with a classical force approach for the 9\,TW/cm$^{2}$ data set: First, the contributing orbitals were calculated using the density functional theory (DFT) code \textsc{FPLO} \cite{koepernik1999full,fplo} for different $d_{SF}$ internuclear distances. Secondly, the impact of the IR pulse was included in a quantum-mechanical many-body restricted active space simulation \cite{pinjari2014restricted} by off-resonant dipole-coupling of the highest occupied molecular orbital (HOMO) and the LUMO. 
The corresponding Hamiltonian is used for the temporal propagation of the electronic wave function of the system via the time-dependent Schr{\"o}dinger equation. 
This results in a laser-dressed ground-state PEC and hence in a slightly changed equilibrium internuclear distance $d_0$. 
Consequently, the nuclei are accelerated classically according to the laser-dressed PEC's slope and result in a new molecular structure after a time step $\delta$t. 
The simulation utilizes the Born-Oppenheimer approximation, where the electronic structure follows the nuclear redistribution immediately. 
Thus, the DFT orbitals of the updated molecular structure are used as input for the next time step. For each time step, the resulting 6a$_{1g}$ absorbance of the SXR probe pulse is calculated with the \textsc{Quanty} code \cite{haverkort2012multiplet,quanty} which results in the black-dashed line in Fig.~3(f). 
The agreement within the SXR-IR temporal pulse overlap is further improved (white solid line) by calculating Stark shifts of the effective LUMO through its dipole-coupling to unoccupied orbitals with ungerade symmetry (LUMO\,+\,\textit{n}) via 
\begin{figure}[h!b]
    \centering
    \includegraphics[width=0.47\textwidth]{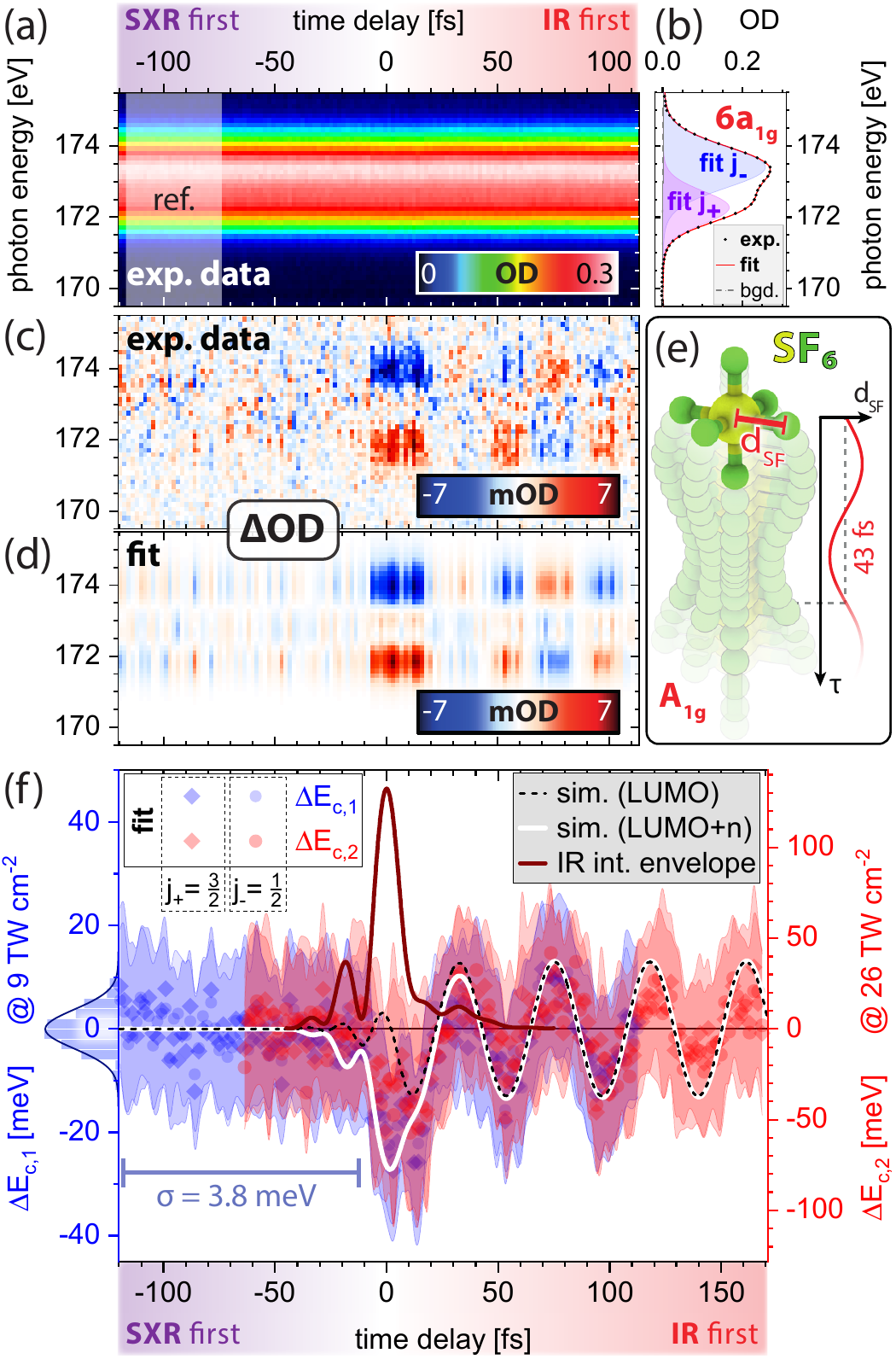}
    \caption{\textbf{TR-XAS data at the sulfur $\boldsymbol{L_{2,3}}$-edge in SF$\mathbf{_6}$.}
    (a) Measured absorbance OD($\omega, \tau$) at an IR intensity of 9\,TW/cm$^2$. 
    (b) Spectral lineout from (a) showing the 6a$_{1g}$ doublet resonance as well as its fit.
    (c) Experimental differential absorbance data $\Delta$OD($\omega, \tau$) calculated from (a) by subtracting the indicated reference mean spectrum [\textit{ref.} in (a)].
    (d) $\Delta$OD($\omega, \tau$) based on the fits of the data in (a).
    (e) Illustration of the symmetric vibrational breathing mode A$_{1g}$ in SF$_6$ which has the same period as the oscillatory signature visible in (c) or (d).
    (f) Resonance energy differences $\Delta E_c$ (with respect to \textit{ref.}) extracted from the Voigt fits of the TR-XAS experimental data over the SXR-IR time delay. Two measurements with different IR intensities are shown (blue: 9\,TW/cm$^2$ and red: 26\,TW/cm$^2$, left and right  ordinate axis, respectively). For each measurement, the 6a$_{1g}$($j_{\pm}$) peak energy differences are indicated as diamonds/circles including the fit uncertainties. Simulation results without (black dashed line) and with (white solid line) taking LUMO\,+\,\textit{n} orbitals with ungerade symmetry into account. A standard deviation of $\sigma = 3.8$\,meV for the SXR-first region of $\Delta E_{c,1}$ (blue histogram) is determined. Furthermore, the measured IR pulse intensity envelope is illustrated (dark red line).
    }
    \label{fig3}
\end{figure}
\begin{figure*}[h!t]
    \centering
    \includegraphics[width=\textwidth]{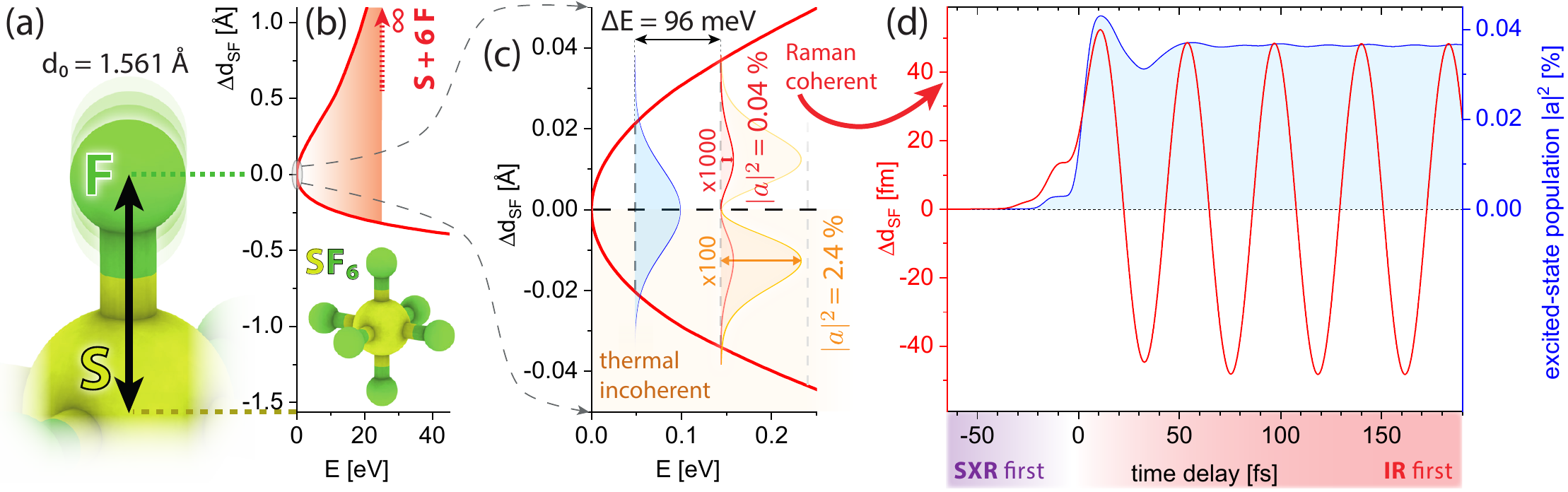}
    \caption{\textbf{Femtometer-resolved, perturbatively-excited vibration in SF$\mathbf{_6}$.}
    (a) Illustration of a single S--F bond in SF$_6$ [full molecular structure in (b)] vibrating around the equilibrium distance $d_0$. (b) Calculated electronic ground-state PEC of symmetric breathing mode A$_{1g}$ vibrations in SF$_6$. (c) Zoom into (b) around $d_0$ with illustrations of the probability density of the vibrational ground state (blue) and the first excited state A$_{1g}$ (red/orange). A comparison is given for the population of A$_{1g}$ due to incoherent thermal excitation at 300\,K (orange) and the 60 times smaller measured coherent ISR excitation (red) by the I$_{IR,1}$~=~9\,TW/cm$^2$ IR pulse. (d) Calculated bond-length difference $\Delta d_{SF}$ and population of A$_{1g}$, $|a|^2$, over SXR-IR time delay. 
    }
    \label{fig4}
\end{figure*}
the measured IR intensity envelope (dark red line).\\
\indent The transient energy shift during time-overlap is due to an electronic AC Stark effect \cite{bakos1977ac}, which is proportional to the IR intensity. 
For time delays $>\,50$\,fs, the observed energy shifts are purely caused by vibrational dynamics. 
Here, the ISR excitation leads to an S--F excursion amplitude which scales linearly with the IR intensity \cite{bartels2003impulsive,shen1965theory}.
Due to the perturbative excitation, the impact of the bond-length change on the resonance energy shift can be approximated to be linear. 
Hence, the overall energy shift $\Delta E_c (t)$ originating from the combined electronic and vibrational dynamics is also proportional to the IR intensity and enables to match the two data sets $\Delta E_{c, 1}$ and $\Delta E_{c, 2}$ in Fig.~3(f) by simply rescaling the ordinate axis by their intensity ratio $I_{IR,1} / I_{IR,2}$. \\
\indent To elucidate the perturbative limit of vibrational excitation, the low IR-intensity (\mbox{I$_{IR, 1}=9$\,TW/cm$^{2}$}) data set $\Delta E_{c, 1}$ is analyzed in more detail: 
For vibrational metrology with TR-XAS, the transition-energy shift 
with respect to the bond-length change is crucial (compare Fig.~1(a) inset). With the calculated electronic many-body ground-state and the many-body excited-state 2t$_{1u}^{-1}$6a$_{1g}^1$ PECs of SF$_6$, a linear slope value of \mbox{$\frac{\Delta E}{\Delta d_{SF}} = 0.27$\,eV/pm} around the ground-state equilibrium S--F distance \mbox{$d_0$ = 1.561\,\AA} \cite{bartell1978structures} is determined. Fig.~4(a) illustrates the symmetric A$_{1g}$ vibration around $d_0$ for a single \mbox{S--F} bond. 
This vibration is governed by the electronic ground-state PEC of the Raman-active A$_{1g}$ vibrations shown in Fig.~4(b). For large vibrational excitations this PEC describes a dissociation to S + 6\,F with a dissociation energy of 22.4\,eV \cite{dibeler1948dissociation}. Due to the perturbative vibrational excitation, only a small region of the PEC around $d_0$ is relevant as illustrated in Fig.~4(c). The vibrational frequency $\nu_{A_{1g}}$~=~775\,cm$^{-1}$ \cite{aboumajd1979analysis,wagner2006monitoring} of the first excited breathing mode A$_{1g}$ in SF$_6$ is equivalent to a vibrational energy-level spacing of 96\,meV. Furthermore, Fig.~4(c) shows the probability densities of the ground- and first-excited-state nuclear wave functions, $\ket{\psi_0}$ and $\ket{\psi_1}$ [compare Fig.~1(b)], respectively, for the case of A$_{1g}$ vibrations in SF$_6$. 
The thermal excitation at room temperature (300\,K) leads to a population of the first excited A$_{1g}$ vibrational state of $|a|^2 = 2.4$\%. In the lower half of Fig.~4(c), the 100 times magnified view of the thermally excited probability density (in orange) is compared to the one of the vibrational ground state (in blue). In contrast, the ISR excitation with a 9\,TW/cm$^{2}$-intensity IR pulse leads to an excited-state population of $|a|^2<0.04$\%, which is illustrated with a magnification of 1000 (in red) in the upper half of Fig.~4(c). Notably, this is a 60 times smaller vibrational excitation compared to the thermally induced population. 
Due to the coherent nature of ISR, however, the excitation in the perturbative limit can be well separated from the incoherent thermal vibrational background. The small deviations from the stepwise behavior of the population dynamics in Fig.~4(d), blue line, are linked to the temporal IR pulse structure. 
As a result, a vibrational amplitude of $\Delta d_{SF} \approx 50$\,fm (compare Fig. 4(d), red line) is extracted from the simulation. 
From the $\sigma=3.8$\,meV standard deviation of the $\Delta E_{c,1}$ data set in Fig.~3(f) within the SXR-first region, a precision of 14\,fm is determined. This equals around two times the 6.5\,fm core diameter of the sulfur atom \cite{schaller1985nuclear} and translates into a relative bond-length-change sensitivity of $10^{-4}$. \\
\indent Overall, our approach for perturbative vibrational control is very general: The widely applicable ISR excitation scheme is easily scalable. Even smallest induced coherent vibrational amplitudes can be separated from the thermal incoherent background and no special preparation of the gas-phase target (e.g., orientation or alignment) is needed. In addition, due to its perturbative nature acting on neutral molecules, the presented technique is potentially of interest for real-time precision observation and control of chemical reactions.
Moreover, our study reveals the technical potential of using an all-optical method for precision metrology of time-dependent molecular dynamics, as well as high-accuracy perturbative quantum control in gas-phase molecules. As the perturbative excitation of a molecular ensemble is measured, there is no fundamental physical restriction concerning exciting and measuring even smaller vibrational amplitudes. Using  a high-resolution SXR spectrometer \cite{kleine2021highly,chiuzbuaian2014design,harada2012ultrahigh} and further optimizations, a sub-femtometer precision seems feasible. 
These results thus substantially advance the field of x-ray spectroscopy in molecules \cite{geneaux2019transient,kraus2018ultrafast,bressler2010molecular} by paving the way for time-resolved vibrational precision metrology.

\begin{acknowledgments}
We thank C. Kaiser and his team for technical support. We acknowledge financial support by the Deutsche Forschungsgemeinschaft (DFG, German Research Foundation) under Germany's Excellence Strategy EXC2181/1-390900948 (the Heidelberg STRUCTURES Excellence Cluster) and by the European Research Council (ERC) (X-MuSiC 616783).
\end{acknowledgments}
\bibliography{lit}
\end{document}